\documentclass[a4paper,11pt]{article}
\usepackage{pos}

\def\zq{${\mathbb Z}_q\ $}              
\def\rar{\rightarrow}                   
\def\beq{\begin{equation}}              
\def\enq{\end{equation}}                

\title{Clock model interpolation and symmetry breaking in O(2) models}
\ShortTitle{Clock model interpolation}

\author*[a,b]{Leon Hostetler}
\author[c]{Jin Zhang}
\author[c]{Ryo Sakai}
\author[d]{Judah Unmuth-Yockey}
\author[b,a]{Alexei Bazavov}
\author[c]{Yannick Meurice}

\affiliation[a]{Department of Physics and Astronomy, Michigan State University, East Lansing, Michigan 48824, USA}

\affiliation[b]{Department of Computational Mathematics, Science and Engineering, Michigan State University, \\East Lansing, Michigan 48824, USA}

\affiliation[c]{Department of Physics and Astronomy, The University of Iowa, Iowa City, Iowa 52242, USA }

\affiliation[d]{Fermilab, Batavia, Illinois 60510, USA}

\emailAdd{hostet22@msu.edu}

\abstract{
The $q$-state clock model is a classical spin model that corresponds to the Ising model when $q=2$ and to the $XY$ model when $q\to\infty$. The integer-$q$ clock model has been studied extensively and has been shown to have a single phase transition when $q=2$,$3$,$4$ and two phase transitions when $q>4$.We define an extended $q$-state clock model that reduces to the ordinary $q$-state clock model when $q$ is an integer and otherwise is a continuous interpolation of the clock model to noninteger $q$. We investigate this class of clock models in 2D using Monte Carlo (MC) and tensor renormalization group (TRG) methods, and we find that the model with noninteger $q$ has a crossover and a second-order phase transition. We also define an extended-$O(2)$ model (with a parameter $\gamma$) that reduces to the $XY$ model when $\gamma=0$ and to the extended $q$-state clock model when $\gamma\to\infty$, and we begin to outline the phase diagram of this model. These models with noninteger $q$ serve as a testbed to study symmetry breaking in situations corresponding to quantum simulators where experimental parameters can be tuned continuously.
}

\FullConference{%
 The 38th International Symposium on Lattice Field Theory, LATTICE2021
  26th-30th July, 2021
  Zoom/Gather@Massachusetts Institute of Technology
}

\begin{document}
\maketitle


\section{Introduction}
\label{sec_intro}

Implementation of quantum field theoretical models on a 
quantum computer (either digital or analog) requires
some form of discretization, either directly in the field
variables or in the expansion of the Boltzmann weight
in the path integral. $q$-state clock models with the
\zq symmetry serve as viable discrete approximations for
models with continuous Abelian symmetries.
To understand the applicability of such discretization
schemes we have recently proposed~\cite{Hostetler:2021uml}
a class of models, called the extended $q$-state clock models,
that interpolates the conventional $q$-state clock models
to noninteger values of $q$. As the $q$-state clock
models reduce to the $O(2)$ model in the limit
$q\to\infty$, we have also constructed a class of
extended-$O(2)$ models where a symmetry breaking term
with a tunable coupling $\gamma$ allows one to smoothly
interpolate between the $O(2)$ model ($\gamma=0$) and
the extended $q$-state clock model 
($\gamma\to\infty$)~\cite{Hostetler:2021uml}.

We present our results on studying these models
in various limits with Monte Carlo (MC) and tensor renormalization
group (TRG) methods and, based on these findings, suggest a 
possible phase diagram of the 
two-dimensional extended-$O(2)$ model
in the $(q,\beta,\gamma)$ parameter space.

\section{A new class of models}
\label{sec_model}

We define the extended-$O(2)$ model by adding a 
symmetry breaking term 
$-\gamma \sum_{x} \cos(q\varphi_x)$ to the action of the
classical $O(2)$ model:
\begin{equation}
	\label{eq_extO2}
	S_{\mbox{\scriptsize ext-}O(2)} = -\beta \sum_{x,\mu}\cos(\varphi_{x+\hat\mu}-\varphi_{x})
	- \gamma\sum_x\cos(q\varphi_x)
	- h \sum_x\cos(\varphi_x-\varphi_h),
\end{equation}
where $\varphi_x$ are angular variables that encode the spin
direction and $(h,\varphi_h)$ are the magnitude and direction
of the external magnetic field.

\begin{figure}
	\centering
		\centering
		\includegraphics[scale=1]{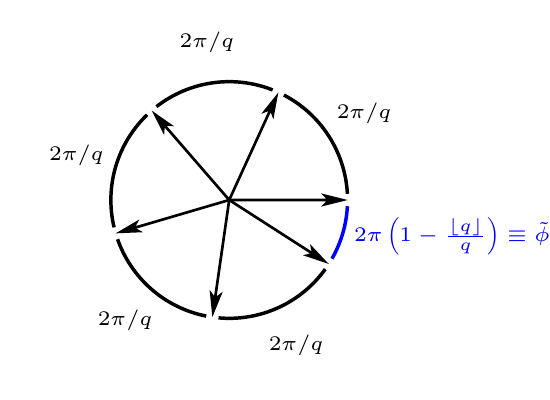}
    \caption{
    Arrows indicate the allowed spin orientations for the extended $q$-state clock model when $q=5.5$. Figure taken from~\cite{Hostetler:2021uml}.\label{allowed_angles_example}
    }
\end{figure}

In the limit $\gamma\to\infty$ the dominant spin directions
are the ones that maximize the second term in 
Eq.~(\ref{eq_extO2}), \textit{i.e.}
$q\varphi_{x}=2\pi k$, $k\in\mathbb{Z}$.
Thus, for integer $q$ this limit reduces the action
in Eq.~(\ref{eq_extO2}) to the one of the $q$-state clock
model, while for noninteger $q$ we consider the 
$\gamma\to\infty$ limit to be the \textit{definition}
of the extended $q$-state clock model where the
spins take the directions
\begin{equation}
\label{eq_qfrac}
0 \leq \varphi^{(k)}_{x}= \frac{2\pi k}{q} <  2\pi.
\end{equation}
These ``allowed'' directions divide the unit circle into $\lceil q\rceil$ arcs of which $\lceil q\rceil-1$ have measure $2\pi/q$ and the small remainder has measure
\begin{equation}
\label{small_angle}
\tilde{\phi} \equiv 2\pi \left(1 - \frac{\lfloor q\rfloor}{q} \right),
\end{equation}
as illustrated in Fig.~\ref{allowed_angles_example}. Here $\lfloor\dots\rfloor$ denotes rounding
down to the nearest integer, and ${\lceil \ldots\rceil}$ denotes rounding up to the nearest integer.

For numerical simulations we consider the extended
$q$-state clock model (\textit{i.e.} $\gamma\to\infty$)
where the angles $\varphi_x$ are allowed to take only the 
discrete values defined in Eq.~(\ref{eq_qfrac}):
\begin{equation}
\label{eq_extq}
S_{\mbox{\scriptsize ext-}q} = -\beta \sum_{x,\mu}\cos(\varphi_{x+\hat\mu}-\varphi_{x})
-h\sum_x\cos(\varphi_x-\varphi_h).
\end{equation}

At noninteger $q$ the action (\ref{eq_extq}) is no
longer invariant under the operation $\varphi\rar\mod(\varphi+2\pi/q,2\pi\lfloor q\rfloor/q)$ (in index notation $k \rightarrow \mod(k+1, \lfloor q \rfloor)$), \textit{i.e.}
the $\mathbb{Z}_q$ symmetry is explicitly broken.
However, there is a residual $\mathbb{Z}_2$ symmetry
with respect to the operation $\varphi\rar2\pi -\varphi-\tilde{\varphi}$ (in index notation 
$k \rightarrow \lfloor q \rfloor - k$).

The partition function is
\begin{equation}
\label{eq_Z}
Z=\sum_{\{\varphi_x\}}e^{-S_{\mbox{\scriptsize ext-}q}},
\end{equation}
and the observables that we compute to study the critical behavior are the internal energy
\begin{equation}
\label{eq_internalenergy}
\langle E\rangle= \langle -\sum_{x,\mu} \cos(\varphi_{x+\hat{\mu}} - \varphi_{x}) \rangle = -\frac{\partial}{\partial\beta}\ln Z,
\end{equation}
the specific heat
\begin{equation}
\label{eq_specificheat}
C= \frac{-\beta^{2}}{V} 
\frac{\partial \langle E \rangle}{\partial \beta} = \frac{\beta^{2}}{V} (\langle E^{2} \rangle - \langle E \rangle ^{2}),
\end{equation}
the magnetization
\begin{equation}
\label{eq_mag}
\langle\vec{M}\rangle=
\frac{\partial}{\partial\vec{h}}\ln Z=
\left\langle
\sum_x\vec{\sigma}_x
\right\rangle,\,\,\,\,\,
\vec{\sigma}_x=(\cos\varphi_x,\sin\varphi_x),
\end{equation}
and the magnetic susceptibility
\begin{equation}
\label{eq_mag_sus}
\chi_{\vec{M}}=\frac{1}{V}\,
\frac{\partial\langle\vec{M}\rangle}{\partial\vec{h}}=
\frac{1}{V}\left(\langle \vec{M} \cdot \vec{M} \rangle-
\langle\vec{M}\rangle \cdot \langle\vec{M}\rangle \right),
\end{equation}
where $V$ is the total number of sites on the lattice.

In a finite system accessible to Monte Carlo simulations
the spontaneous magnetization defined in
Eq.~(\ref{eq_mag}) averages out to 0 if there is no
external magnetic field. For this reason we
use proxy magnetization observables
\begin{equation}
\label{eq_proxymag}
\langle|\vec{M}|\rangle=
\left\langle\left|
\sum_x\vec{\sigma}_x
\right|\right\rangle
\,\,\,\,\,\,\,\,
\mbox{and}
\,\,\,\,\,\,\,\,
\chi_{|\vec{M}|}=
\frac{1}{V}\left(\langle |\vec{M}|^2\rangle-
\langle|\vec{M}|\rangle^2\right),
\end{equation}
as is often done in Monte Carlo simulations,
\textit{e.g.} Ref.~\cite{Murty:1984}.

\section{Monte Carlo Results}
\label{sec_results}

We performed Monte Carlo simulations for the
two-dimensional extended $q$-state clock model
without an external magnetic field on $4\times4$
lattice. 
The updating was performed with 
a version of the heatbath
algorithm from Ref.~\cite{berg:2004}
modified appropriately to handle noninteger values of $q$.
In this initial round we mainly focused on scanning
the parameter space and did not pursue larger
lattices with Monte Carlo since the local updating
algorithms suffer significant slowing down in the 
ordered phase at large values of the inverse temperature
$\beta$.

Typically, we started from a random configuration
(hot start)
and performed $2^{15}$ sweeps for equilibration followed
by a production run with $2^{30}$ sweeps. The measurements
were performed once in $2^8$ sweeps, giving us a time
series of length of $2^{22}$ for averaging and error analysis.
For error propagation we used the jackknife method
with $2^6$ jackknife bins as described in
Ref.~\cite{berg:2004}.

\begin{figure*}
	\includegraphics[width=\textwidth]{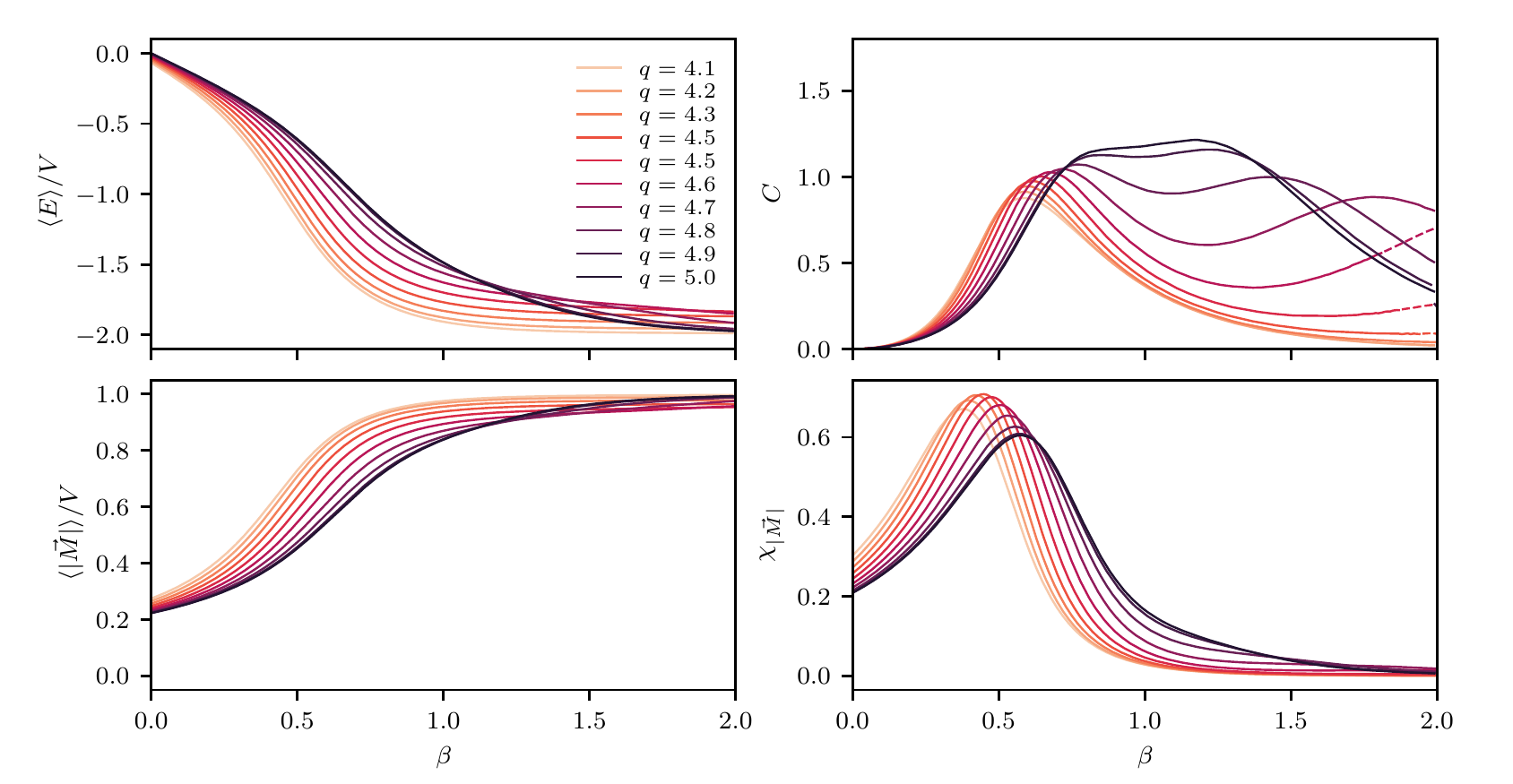}
	\caption{
	Monte Carlo results from a heatbath algorithm for the extended $q$-state clock model on a $4\times 4$ lattice with $4<q\leq 5$. The four panels show the energy density $\langle E\rangle/V$, specific heat $C$, proxy magnetization density $\langle |\vec{M}|\rangle/V$ and magnetic susceptiblity $\chi_{|\vec{M}|}$. Statistical error bars are omitted since they are smaller than the line thickness. Dashed lines, where they occur, indicate regions where we have data but the uncertainty is not fully under control. Plots taken from~\cite{Hostetler:2021uml}.\label{zz_fours_all_0004_0p00_23_1_encv}}
\end{figure*}

Results for $4.1 \leq q \leq 5.0$ are shown in Fig.~\ref{zz_fours_all_0004_0p00_23_1_encv}. The four panels show the energy density and the specific heat defined in Eqs.~(\ref{eq_internalenergy}) and~(\ref{eq_specificheat}) and the proxy magnetization and susceptibility defined in Eq.~(\ref{eq_proxymag}). We see a double-peak structure in the specific heat. As $q\rightarrow 4$ from above, the peak at large-$\beta$ moves toward $\beta=\infty$. As $q\rightarrow 5$ from below, the specific heat becomes that of the ordinary 5-state clock model. In the susceptibility, a double-peak structure appears at larger lattice sizes. In general, we find that the thermodynamic curves vary smoothly for $n<q\leq n+1$ where $n$ is an integer. When $q$ goes from $n$ to $n+\epsilon$, the thermodynamic curves change abruptly since the number of spin orientations is increased by one. At $q = n+\epsilon$, the specific heat exhibits a double-peak structure with the second peak at very large $\beta$. As $q$ is increased further, this second peak moves toward small $\beta$, until at $q=n+1$, the thermodynamic curves of the integer-$(n+1)$-state clock model are recovered.

In the small-$\beta$ (high temperature) regime, all allowed angles of the extended $q$-state clock model are essentially equally accessible. In this regime, the model is dominated by the approximate $\mathbb{Z}_{\lceil q\rceil}$ symmetry, and it behaves approximately like a ${\lceil q\rceil}$-state clock model. In the large-$\beta$ (low temperature) regime, the model is dominated by the residual $\mathbb{Z}_2$ symmetry, and it becomes a rescaled Ising model.

\begin{figure}%
	\centering
		\includegraphics[scale=1]{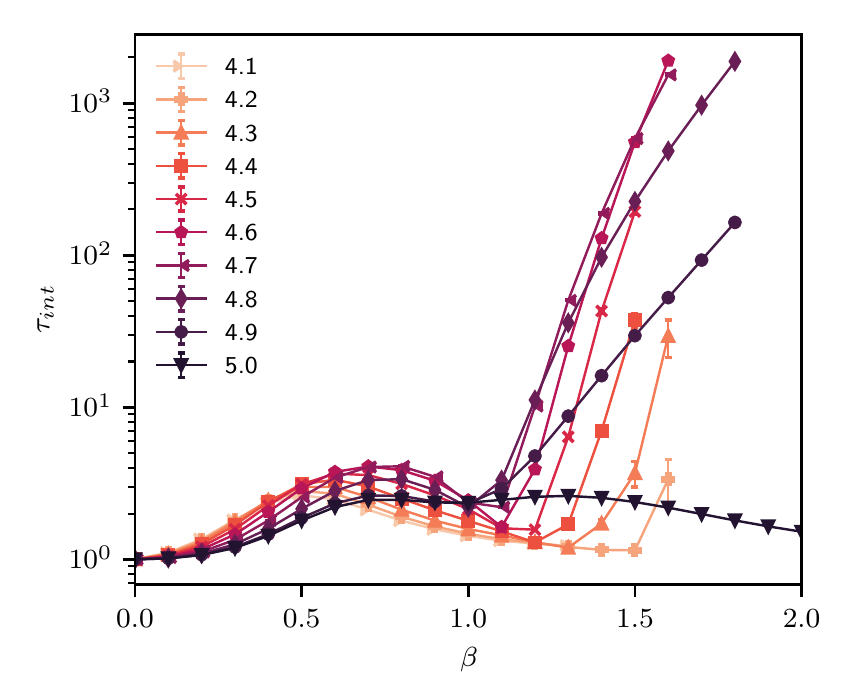}
    \caption{The integrated autocorrelation time $\tau_{int}$ (defined in Eq.~(\ref{eq_ace})) in the extended $q$-state clock model for $4< q\leq 5$ on a $4 \times 4$ lattice using a heatbath algorithm. Here, the observable used is the energy density. For integer $q$, $\tau_{int}$ is well-behaved and grows moderately near the critical points. For noninteger $q$, $\tau_{int}$ grows rapidly at large $\beta$ resulting in a critical slowing down of the Monte Carlo method. Note the log scale on the vertical axis. Connecting lines are included to guide the eyes. Plot taken from~\cite{Hostetler:2021uml}. \label{zz_all_all_0004_0p00_var_1_ac2}}
\end{figure}

At large-$\beta$, all the spins tend to magnetize along one specific direction among the ${\lceil q\rceil}$ allowed directions. However, when $q\notin\mathbb{Z}$, the $\mathbb{Z}_{\lceil q\rceil}$ symmetry is broken, and these directions are not all equivalent. To understand this, we can consider the case $q = 5.5$ with the allowed spin directions illustrated in Fig. \ref{allowed_angles_example}. For example, a configuration magnetized in the direction $2\pi/q$ is not equivalent to one magnetized in the direction 0. In the first case, the probability of the spins flipping to a new direction is low because the neighboring directions are far away. In the second case, the probability of the spins flipping to a new direction is relatively high because of the small angular distance between directions $0$ and $-\tilde{\phi}$. Thus, we find that at large $\beta$, the configuration space separates into two thermodynamically distinct sectors, and the Markov chain has trouble adequately sampling both sectors. The resulting Monte Carlo slowdown is illustrated by a sudden increase of the integrated autocorrelation time in the intermediate-$\beta$ regime as shown in Fig.~\ref{zz_all_all_0004_0p00_var_1_ac2}.

For an observable $O$, an estimator of the integrated autocorrelation time is given by \cite{berg:2004}
\beq
\label{eq_ace}
\tilde{\tau}_{O,int}(T) = 1+ 2\sum_{t=1}^T \frac{C(t)}{C(0)},
\enq
where $C(t) = \langle O_i O_{i+t}\rangle - \langle O_i\rangle \langle O_{i+t}\rangle$ is the correlation function between the observable $O$ measured at Markov times $i$ and $i+t$. The integrated autocorrelation time $\tau_{O,int}$ is estimated by finding a window in $T$ for which $\tilde{\tau}_{O,int}(T)$ is nearly independent of $T$.

To appropriately sample the configuration space we had to use large statistics. The effect of autocorrelation in our results was mitigated by discarding $2^8$ heatbath sweeps between each saved measurement. The saved measurements were then binned (i.e. preaveraged) with bin size $2^{16}$ before calculating the means and variances. This approach was adequate for the $4\times 4$ lattice, but the Monte Carlo slowdown makes it difficult to study larger lattices, and is a strong motivation for using TRG.

\section{TRG Results}

We studied the extended $q$-state clock model using a tensor renormalization group (TRG) method. This method does not suffer from the slowdown experienced with the Monte Carlo method, and so the model could be studied on much larger lattices and in the thermodynamic limit, allowing us to perform finite-size scaling and to characterize the phase transitions. The TRG results were validated by comparison with exact and Monte Carlo results on small lattices. A full description of the method is given in \cite{Hostetler:2021uml}.

\begin{figure*}
	\centering
	\includegraphics[width=\textwidth]{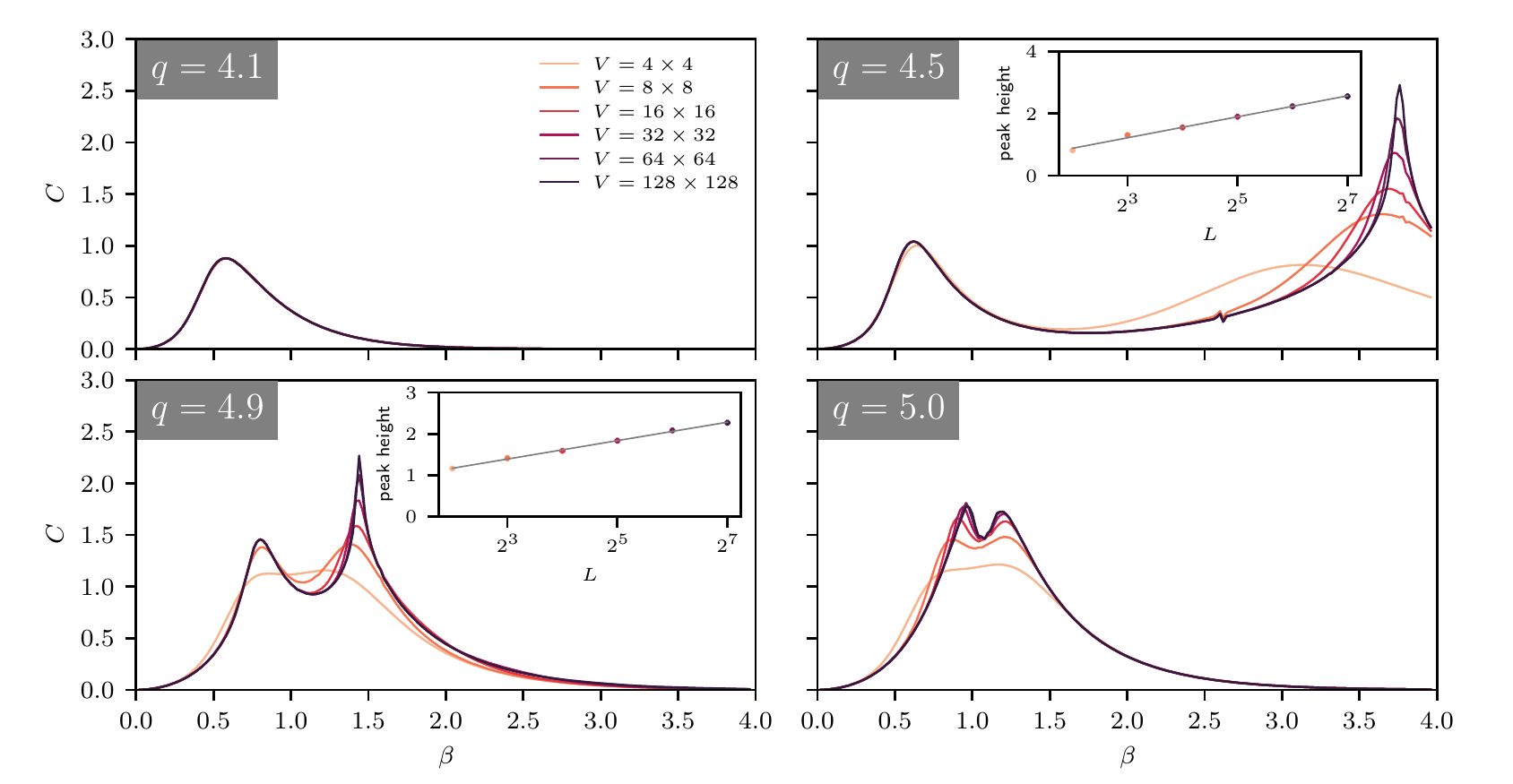}
	\caption{
	The specific heat of the extended $q$-state clock model from TRG for $q=4.1$, 4.5, 4.9, and 5.0. Each panel shows results for lattice sizes ranging from $4\times 4$ to $128\times 128$. In all four examples, there is a double-peak structure (for $q=4.1$ the second peak is at $\beta\sim 75$). Insets show that the second peak grows logarithmically with the linear system size $L=\sqrt{V}$ when $q$ is noninteger. Plots taken from~\cite{Hostetler:2021uml}.
	}
	\label{3figs}
\end{figure*}

In Fig.~\ref{3figs}, we show the specific heat from TRG for $q = 4.1$, 4.5, 4.9, and 5.0 at volumes ranging from $4\times 4$ to $128\times 128$. In general, there are two peaks in the specific heat. When $q=5.0$, the two peaks show little dependence on volume for lattice sizes larger than $32 \times 32$. This is consistent with the two BKT transitions \cite{liping2020} of the ordinary 5-state clock model. When $q\notin\mathbb{Z}$, we see that the first peak shows similar behavior, which indicates that the first peak is associated with either a crossover or a phase transition with an order larger than two. In contrast, the second peak grows logarithmically with volume, as shown in the insets for $q = 4.5, 4.9$. This indicates that the second peak is associated with a second-order phase transition. However, to conclusively characterize the phase transitions associated with these two peaks in the extended $q$-state clock model, we study the magnetic susceptibility with a weak external field in the thermodynamic limit.

\begin{figure}
    \centering
    \begin{minipage}{0.49\textwidth}
        \centering
        \includegraphics[width=0.9\textwidth]{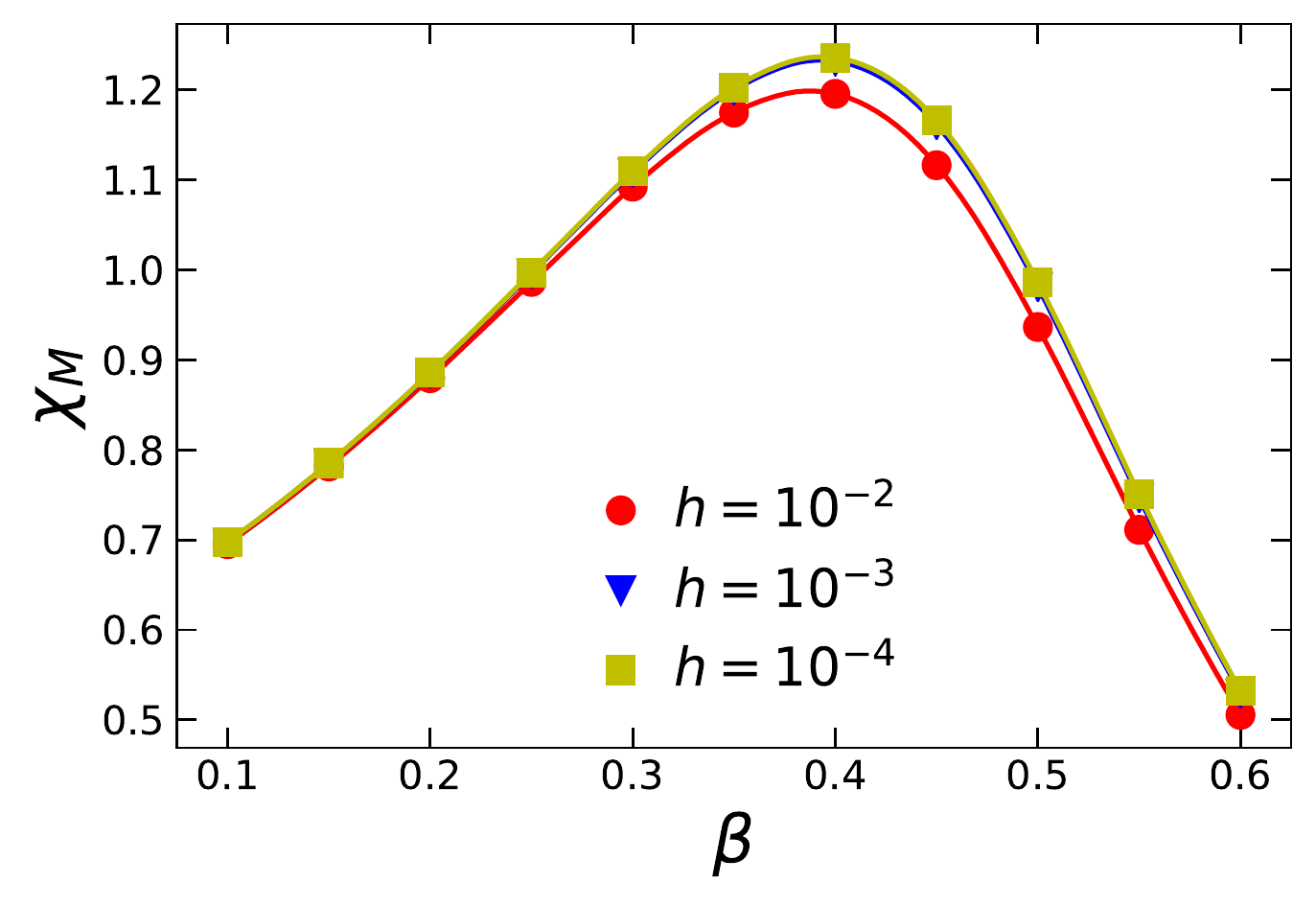}
        \caption{\label{fig:q4p3msus1stpeak} The small-$\beta$ peak of the magnetic susceptibility as a function of $\beta$ for different values of the external magnetic field $h$. The peak height converges to a finite constant value when the external field is taken to zero, indicating that there is no phase transition associated with this peak. In this example, $q = 4.3$. Plot taken from~\cite{Hostetler:2021uml}.}
    \end{minipage}\hfill
    \begin{minipage}{0.49\textwidth}
        \centering
        \includegraphics[width=0.95\textwidth]{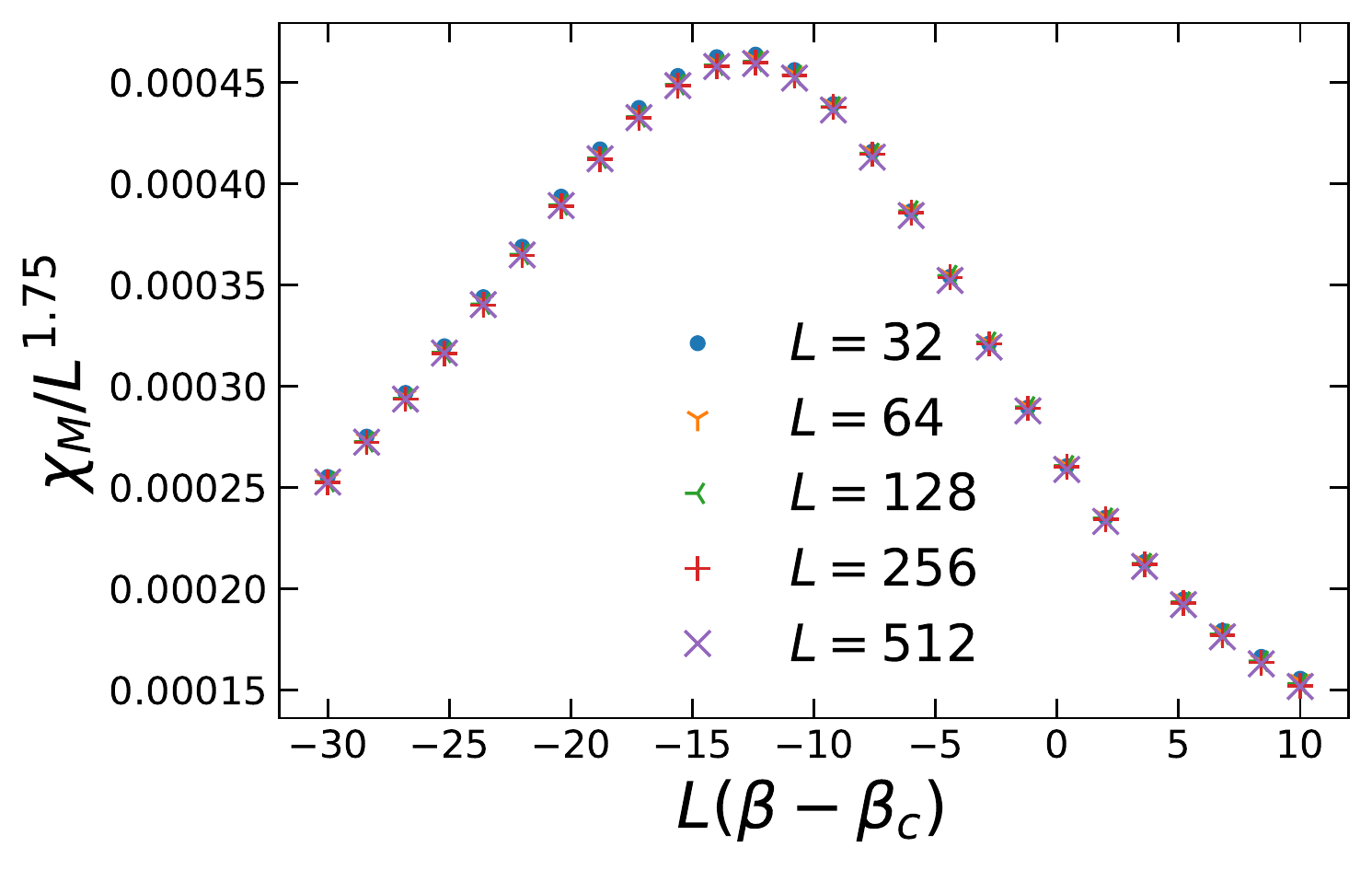}
        \caption{\label{fig:q4p3msuscollapse}We plot $\chi_M/L^{1.75}$ versus $L(\beta-\beta_c)$ in the vicinity of the large-$\beta$ peak for different lattice sizes. The curves collapse to a single universal curve providing strong evidence that this peak is associated with a critical point of the Ising universality class. In this example, $q = 4.3$, we fix $hL^{15/8} = 40$, and $\beta_{c} \approx 9.3216$ is given by Eq.~(\ref{eq:isingcritical}). Plot taken from~\cite{Hostetler:2021uml}.}
    \end{minipage}
\end{figure}

In Fig.~\ref{fig:q4p3msus1stpeak}, we show an example of the magnetic susceptibility in the thermodynamic limit and in the vicinity of the first peak for $q=4.3$ and several different values of the external magnetic field. As the external field is decreased, the peak height converges to a constant $\chi_M \approx 1.2$. Since $\chi_M$ does not diverge as $h\rightarrow 0$, there is no phase transition associated with this peak. This is true for all fractional $q$, and so for fractional $q$, the first peak in the specific heat is associated with a crossover rather than a true phase transition.

Whereas the first peak in the magnetic susceptibility converges to a constant value when the external field $h$ is taken to zero, the second peak diverges. When the peak heights $\chi_M^*$ of the magnetic susceptibility for small values of $h$ are plotted and a power-law extrapolation to $h=0$ is performed, we find that $\chi_M^* \sim h^{-0.93318(16)}$. This gives a magnetic critical exponent $\delta = 14.97(4)$, which is consistent with the value $\delta=15$ associated with BKT and Ising transitions. However, a BKT transition should be accompanied with a continuous critical region, so the divergent large-$\beta$ peak of $\chi_M$ must be an Ising critical point. In the Ising universality class, $\chi_M \sim | \beta - \beta_c |^{-\gamma_e} \sim L^{7/4}$, where $\gamma_e = 7/4= 1.75$ is a universal critical exponent. There is a universal function relating $\chi_M / L^{1.75}$ and $L(\beta-\beta_c)$ with fixed $hL^{15/8}$. In Fig.~\ref{fig:q4p3msuscollapse}, we plot $\chi_M/L^{1.75}$ versus $L(\beta-\beta_{c})$ for various lattice sizes around the large-$\beta$ peak of $\chi_M$ for $q=4.3$. We see that the data collapse to a single curve, and this is strong evidence that the large-$\beta$ peak is a critical point of the Ising universality class. Here, the critical point $\beta_c$ of the extended $q$-state clock model is approximated by the critical point of a rescaled Ising model,
\begin{eqnarray}
\label{eq:isingcritical}
\beta_c \simeq \frac{\ln\left(1+\sqrt{2}\right)}{1-\cos{\tilde{\phi}}}.
\end{eqnarray}
The small angular distance $\tilde{\phi}$ depends on $q$ and is defined in Eq.~(\ref{small_angle}).

\section{Phase diagram}
\label{sec_results_phasediag}

\begin{figure}
\centering
\begin{minipage}{.5\textwidth}
\includegraphics[width=\textwidth]{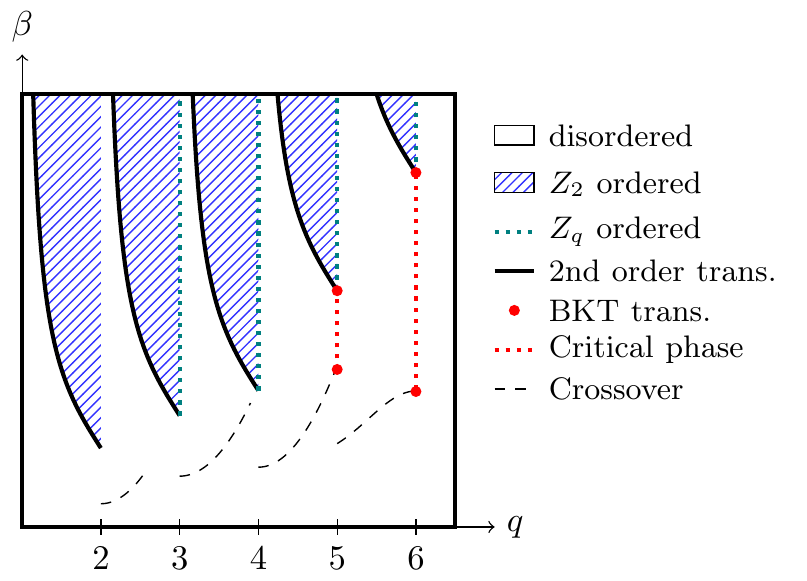}
\end{minipage}\hfill
\begin{minipage}{.38\textwidth}
\includegraphics[width=\textwidth]{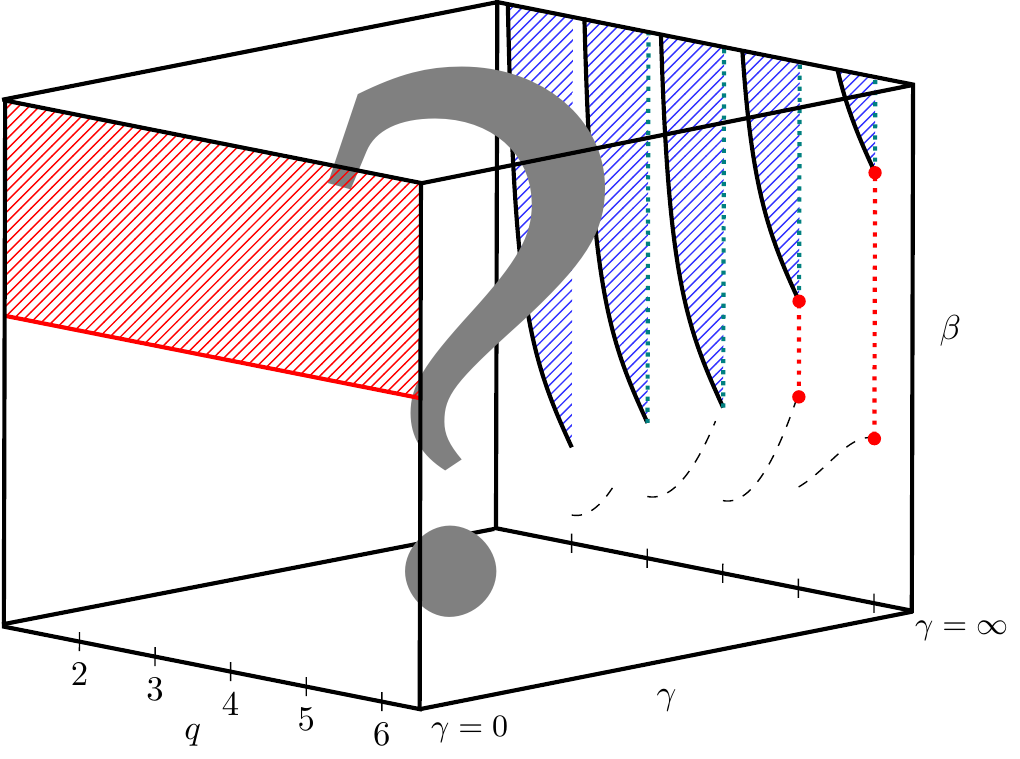}
\end{minipage}
\caption{
\textbf{(Left)} The phase diagram of the extended $q$-state clock model as determined by a TRG study of the magnetic susceptibility. For integer $q$, the model reduces to the ordinary clock model, which has a second-order phase transition when $q=2,3,4$ and two BKT transitions when $q\geq 5$. For noninteger $q$, there is a crossover at small $\beta$, and a second-order phase transition at larger $\beta$. \textbf{(Right)} We outline the 3-dimensional phase diagram of the extended-$O$(2) model. When $\gamma=0$, the model reduces to the $XY$ model for all values of $q$. The $XY$ model has a single BKT transition separating a disordered phase at small $\beta$ from a critical phase at large $\beta$. In the $\gamma=\infty$ plane, the extended-$O$(2) model reduces to the extended $q$-state clock model. The phase structure at finite-$\gamma$ will be addressed in future work. Figures taken from~\cite{Hostetler:2021uml}.\label{phasediagrams}
    }
\end{figure}

For integer $q$, the extended $q$-state clock model reduces to the well-known ordinary clock model \cite{Ortiz:2012}. For $q=2,3,4$, there is a disordered phase and a $\mathbb{Z}_q$ symmetry-breaking phase separated by a second-order phase transition. For $q \geq 5$, there are two BKT transitions \cite{liping2020}. There is a disordered phase at small-$\beta$, and a $\mathbb{Z}_q$ symmetry-breaking phase at large-$\beta$ with a critical phase at intermediate $\beta$. For noninteger $q$, the extended $q$-state clock model has a double-peak structure in both the specific heat and the magnetic susceptibility. Using TRG, we have shown that the small-$\beta$ peak is associated with a crossover, and the large-$\beta$ peak is associated with a phase transition of the Ising universality class. Thus, we get the phase diagram shown in the left panel of Fig.~\ref{phasediagrams}.

In the extended-$O$(2) model, with action given by Eq.~(\ref{eq_extO2}), the phase diagram is three-dimensional with parameters $\beta$, $q$, and $\gamma$. The $\gamma=0$ plane of this model is the $XY$ model for all values of $q$, and the $\gamma=\infty$ plane is the extended $q$-state clock model. Thus, an outline of the phase diagram of this model is given in the right panel of Fig.~\ref{phasediagrams}. In future work, we will study this model with finite $\gamma$.

\section{Summary and Outlook}
\label{sec_summary}

We defined an extended-$O$(2) model by adding a symmetry breaking term $\gamma \cos(q\varphi_x)$ to the action of the two-dimensional $O(2)$ model. In the $\gamma\rar\infty$ limit, the spins are forced into the directions $0 \leq \varphi^{(k)}_{x}= 2\pi k/q <  2\pi$ with $k\in\mathbb{Z}$. We take this limit as the definition of the extended $q$-state clock model since in this limit, when $q$ is integer, the ordinary $q$-state clock model is recovered. In this work, we studied the extended $q$-state clock model for noninteger $q$ using Monte Carlo and tensor renormalization group methods. We found that there are two peaks in both the specific heat and the magnetic susceptibility. The small-$\beta$ peak is associated with a crossover, and the large-$\beta$ peak is associated with an Ising phase transition. Thus we obtained the phase diagram of the extended $q$-state clock model and began to outline the 3-parameter phase diagram of the extended-$O$(2) model. The full phase diagram of the extended-$O$(2) model will be discussed in future work.

Interpolations among $\mathbb{Z}_n$ clock models have been realized experimentally using a simple Rydberg simulator \cite{keesling2019}. In this experimental work, $\mathbb{Z}_n$ ($n \ge 2$) symmetries emerge by tuning continuous parameters---the detuning and Rabi frequency of the laser coupling, and the interaction strength between Rydberg atoms. Such work paves the way to quantum simulation of lattice field theory with discretized field variables. Several concrete proposals for discretizing fields using similar configurable arrays of Rydberg atoms were recently put forward \cite{meurice2021}.

\acknowledgments

We thank Gerardo Ortiz, James Osborne, Nouman Butt, Richard Brower, and  members of the QuLAT collaboration for useful discussions and comments. This work was supported in part by the U.S. Department of Energy (DOE) under Awards No. DE-SC0010113 and No. DE-SC0019139. The Monte Carlo simulations were performed at the Institute for Cyber-Enabled Research (ICER) at Michigan State University.


\providecommand{\href}[2]{#2}\begingroup\raggedright\endgroup

\end{document}